\documentclass[12pt]{article}
\usepackage{amssymb,latexsym}
\usepackage[cmex10]{amsmath}
\usepackage[all]{xy}
\input{xy}
\xyoption{all}
\title{On the Annihilator Ideal  of an Inverse Form. Simplificandum.}
\author{Graham H. Norton
\footnote{School of Mathematics and Physics, University of Queensland, Brisbane, Queensland 4072, Australia. (Email: ghn@maths.uq.edu.au)}
}

\newtheorem{theorem}{\bf Theorem}[section]
\newtheorem{corollary}[theorem]{\bf Corollary}
\newtheorem{proposition}[theorem]{\bf Proposition} 
\newtheorem{notation}[theorem]{\sc Notation} 
\newtheorem{definition}[theorem]{\bf Definition} 
\newtheorem{lemma}[theorem]{\bf Lemma} 
\newtheorem{example}[theorem]{\it Example}  
\newtheorem{examples}[theorem]{\it Examples}  
\newtheorem{algorithm}[theorem]{\bf Algorithm} 
\newtheorem{conjecture}[theorem]{\bf  Conjecture}
\newtheorem{remark}[theorem]{\it Remark}
\newtheorem{remarks}[theorem]{\it  Remarks}

\newenvironment{proof}{{\noindent\it Proof.\ }}{$\square$\par\vspace{4mm}} 
\def \bt{ \begin{theorem} }
\def \et{ \end  {theorem} }
\def \bl{ \begin{lemma} }
\def \el{ \end  {lemma} }
\def \bp{ \begin{proposition} }
\def \ep{ \end  {proposition} }
\def \bn{ \begin{notation} }
\def \en{ \end  {notation} }
\def \bq{ \begin {question} }
\def \eq{ \end {question} }
\def \bc{ \begin{corollary} }
\def \ec{ \end  {corollary} }
\def \bcj{ \begin{conjecture} }
\def \ecj{ \end  {conjecture} }
\def \bd{ \begin{definition} }
\def \ed{ \end  {definition} }
\def \bdp{ \begin{definitionprop} }
\def \edp{ \end  {definitionprop} }
\def \bdt{ \begin{definitiontheorem} }
\def \edt{ \end  {definitiontheorem} }
\def \bpr{ \begin  {proof} }
\def \epr{ \end  {proof} }
\def \ba{ \begin{algorithm} }
\def \ea{ \end{algorithm} }
\def \be{ \begin{example} }
\def \eex{ \end{example} }
\def \bes{ \begin{examples} }
\def \eexs{ \end{examples} }
\def \br{ \begin{remark} }
\def \er{ \end{remark} }
\def \brs{ \begin{remarks} }
\def \ers{ \end{remarks} }
\def \bpb{ \begin{problem} }
\def \epb{ \end{problem} }

\newcommand{\E}{\mathrm{E}}

\newcommand{ \Gb} {Groebner basis\ }

\newcommand{\la } {\leftarrow}

\newcommand{\ol} {\overline}

\newcommand{\ul} {\underline}
\newcommand{\ee} {\mathrm{E}}

\newcommand{\m} {\mathfrak{m}}

\newcommand{\sm} {\,\sharp\,}

\newcommand{\rem} {\mathrm{rem\,}}

\newcommand{\vv} {\mathrm{v}}

\newcommand{\lc} {\mathrm{LC}}
\newcommand{\LC} {\ell}

\newcommand{\LM} {\mathrm{LM}}
\newcommand{\LT} {\mathrm{LT}}
\newcommand{\N} {\mathbb{N}}
\newcommand{\K} {\mathbb{K}}

\newcommand{\Zdiv}{\mathrm{Zdiv}}

\newcommand{\R} {\mathrm{R}}

\newcommand{\calE}{\mathcal{E}}
\newcommand{\F}{\mathcal{F}}

\newcommand{\Z}{\mathbb{Z}}

\newcommand{\D} {\mathcal{D}}
\newcommand{ \G} {\mathcal{G}}
\newcommand{\I} {\mathcal{I}}
\newcommand{\J} {\mathcal{J}}

\newcommand{\M} {\mathrm{M}}

\begin{document}
\maketitle
\begin{abstract}
We simplify an earlier paper of the same title by not using syzygy polynomials and  by not using a trichotomy of inverse forms.
Let $\K$ be a field and $\M=\K[x^{-1},z^{-1}]$ denote Macaulay's $\K[x,z]$ module of inverse polynomials; here   $z$ and $z^{-1}$ are   homogenising variables. 
An inverse form  $F\in\M$  has a homogeneous annihilator ideal, $\I_F$\,.   In an earlier paper we inductively constructed  an  ordered pair ($f_1$\,,\,$f_2$)  of forms in $\K[x,z]$ which generate $\I_F$. We  used syzygy polynomials to show that the intermediate forms give a minimal grlex Groebner basis, which can be efficiently reduced.
 
We give a significantly shorter proof  that the  intermediate forms are a minimal grlex Groebner basis for $\I_F$\,.    We also simplify our proof that either $
F$ is already reduced or a monomial  of $f_1$ can be reduced by  $f_2$\,. 
The algorithm that computes $f_1\,,f_2$ yields a variant of the Berlekamp-Massey algorithm  which does not use the last 'length change' approach of Massey.
 These new proofs avoid the three separate cases, 'triples' and the technical factorisation of intermediate 'essential' forms.
We also show that $f_1,f_2$ is a maximal  $\R$ regular sequence for $\I_F$\,, so that $\I_F$ is a complete intersection. 
\end{abstract}

{\small {\bf Keywords:} 
 Annihilator ideal,   Berlekamp-Massey algorithm, grlex, Groebner basis,   inverse form, regular sequence.}\\
\section{Introduction}
\subsection{Background}
Let $\K$ be a field. In  \cite[Section 2]{AD},  the authors 
 defined a 'generating form' for a finite sequence $s$ using  negative Laurent  series in two variables, considered as a $\K[x,z]$ module. It is  based on  Macaulay's inverse system.  They showed that their annihilator ideal is homogeneous and hence is generated by finitely many forms. In   \cite[Section 4]{AD} the authors gave an extension of  Berlekamp-Massey  (BM) algorithm and showed that it yields  
 a  minimal homogeneous graded-lexicographic  \Gb  for their annihilator ideal. 
They also showed that if a sequence $s$ has $n\geq 1$ terms,
 linear complexity $\LC_s$ and $2\LC_s\leq n$ then their \Gb for the annihilator ideal is its unique  reduced grlex Groebner basis, \cite[Lemma 7]{AD}. 

In \cite{N15b} we simplified and extended \cite[Sections 2,4]{AD}. We used the $\K[x,z]$ module of inverse polynomials $\K[x^{-1},z^{-1}]$ of inverse polynomials due to Macaulay.  We  defined the  annihilator ideal $\I_F$ of a non-zero inverse form $F$, which is homogeneous. We inductively worked with a special 'viable' ordered pair $f$ of  forms which generate $\I_F$\,.
This gave an effective Hilbert Basis theorem for $\I_F$\,. 
 Accumulating  intermediate forms gives a tuple $\F$ headed by our viable ordered pair and length at most $\lambda_F+1$; $\lambda_F$ is our analogue of $\lambda_s$. We derived a syzygy for our pair and inductively applied Buchberger's syzygy polynomial criterion to show that $\F$ is a minimal grlex \Gb of $\I_F$\,. We also showed how to obtain the reduced \Gb of $\I_F$ directly from $\F$.
 
Section 2 contains some algebraic preliminaries and in Section 3 we review the necessary results from \cite{N15b}. Then we give a short  direct proof that $\F$ is a minimal \Gb for $\I_F$. This new proof does not use the syzygy polynomials of Buchberger as in \cite{N15b}.  In particular, we do not use the technical notion of an essential inverse form (i.e. $\lambda_F>\lambda_0$) nor do we have to treat the cases $F=1$, $F$ geometric and $F$ essential separately. The same applies  to obtaining the reduced \Gb of $\I_F$. Algorithm 5.24 of \cite{N15b} still applies and is an example of computing a reduced \Gb without using Buchberger's algorithm.

In Section 4, we simplify some additional results from \cite{N15b}; the new proofs likewise no longer require separate cases, factorisation of $f_2$ and the technical notion of a 'triple'. We  do not use the last 'length change' of  \cite[Formula (11)]{Ma69}. (We were led to incorporate the trichotomy as the last 'length change' is not defined if $s=1$ or $s$ is geometric.)

In the final section we show that  $\I_F$ is $\m$-primary and that our generators $f_1,f_2$ give a maximal $\R$ regular sequence for $\I_F$. In particular, $\I_F$ is a complete intersection.
\section{Algebraic Preliminaries}
\subsection{Notation}
We continue the notation and conventions from \cite{N15b}. 
 For any set $S$ containing 0, $S^\times=S\setminus\{0\}$ so that $\N^\times=\{1,2,\ldots\}$.  Throughout the paper, $\K$ is an arbitrary field and $\R=\K[x,z]$.  Multiplication in $\R$ is written as juxtaposition. For $\varphi,\varphi'\in\R$ and $k\in\N^\times$,  $x^k\,\varphi+\varphi'$ means $x^k\varphi(x,z)+\varphi'(x,z)$ and similarly for  $\varphi+\varphi'\,z^k$\,; for $p\in\N$, $z^p||\varphi$ if $z^p|\varphi$ but $z^{p+1}\nmid \varphi$.  The total degree of $\varphi\in\R^\times$ is $|\varphi|$,  with $|x|=|z|=1$. 
 The ideal  generated by $\varphi_1,\ldots ,\varphi_m\in\R$ is written $\langle \varphi_1,\ldots ,\varphi_m\rangle$ and $\m$ is the maximal ideal $\langle x,z\rangle$ of $\R$.

We also include reference tables of commonly-used symbols for the aid of the reader.

\begin{center}
\begin{tabular}{|c|l|}\hline
Symbol & Meaning \\\hline\hline
$a\sm F$& the inverse form $a\,x^{m-1}+Fz^{-1}$, $F\neq0$ \\\hline
$d=d_f$ &$|f_2|-|f_1|$\\\hline
$\D$ & a tuple   with $z^{\D_i-m}||f_i$\\\hline
$f=(f_1,f_2)$ & the constructed viable ordered pair for $\I_F$\\\hline
$f_i$ & elements of $\F$\\\hline
$F$ & a (non-zero) inverse form in $\M^\times$\\\hline
$ F^{(i)}$& the 'subform' $F_i\,x^i+\cdots+x^\vv\,z^{i-\vv}$ of $F$, $m\leq i\leq\vv$\\\hline
$\F$ &  the constructed form tuple  for $\I_F$\\\hline
$|\F|$ &   $\F$ and $\D$ are $|\F|$-tuples  \\\hline
$\I_F$ & the annihilator ideal of $F$\\\hline
$m$&  the non-positive total degree of $F$\\\hline
$q=q_f$& $\Delta(f_1;a\sm F)/\Delta(f_2;a\sm F)$\\\hline
$\vv=\vv(F)$ & the order of $F$.\\\hline
\end{tabular}
\end{center}
\begin{center}
\begin{tabular}{|c|l|}\hline
Greek Symbol & Meaning \\\hline\hline
$\Delta_i=\Delta(f_i;a\sm F)$ &  the discrepancy of  $f_i$ and $a\sm F$, $i=1,2$\\\hline
$\lambda_F$ & a non-negative integer derived from $\I_F\cap \Phi$\\\hline
$\varphi$ & a form in $\R^\times$\\\hline
$\varphi\circ F$ & $\varphi$ acting on $F$\\\hline
$\Phi$ & the set of non-zero, monic forms $\varphi$ with $z\nmid\LM(\varphi)$.\\\hline
\end{tabular}
\end{center}

\subsection{Grlex}
We adopt \cite{IVA} as a general reference. We write $\succ$ for graded-lexicographic order ({\em grlex}) on monomials of $\R^\times$, with $|x|=|z|=1$ and $x\succ z\succ1$. Recall that  $\succ$ is the linear  order defined on monomials of $\R^\times$ as follows: $M\succ M'$ if $|M|>|M'|$ or 
($|M|=|M'|$ and $M>_{\mathrm{lex}}M'$). 
 We write $\ee(\varphi)$ for the  grlex {\em exponent} or multidegree of $\varphi\in\R^\times$ :
 $$\ee(\varphi)=\max_\succ\{i\in\N^2:\varphi_i\neq 0\}$$
 and $\LM(\varphi)$ is its leading  monomial; the leading coefficient of $\varphi$ is $\lc(\varphi)$ and its leading term is $\LT(\varphi)=\lc(\varphi)\LM(\varphi)$.
 If   $\varphi$ is also a   form and $d=|\varphi|\in\N$, it will be convenient to write  $\varphi=\sum_{j=0}^d\varphi_j\,x^j\,z^{d-j}$ on the understanding that $\varphi_j=0$ for $(j,d-j)\succ \ee(\varphi)$.  In this case  $z|\varphi$ if and only if $z|\LM(\varphi)$ and if $z\nmid \LM(\varphi)$ then $\ee(\varphi)=(|\varphi|,0)$.
 \subsection{Inverse Forms}

 We also order the  monomials of $\M^\times=\K[x^{-1},z^{-1}]^\times$ using grlex, now written $\prec$\,, but with $|x^{-1}|=|z^{-1}|=-1$, $x^{-1}\prec z^{-1}\prec1$ and
  $$\ee(F)=\min_\prec\{i\in-\N^2:F_i\neq 0\}$$
  is the $\prec$ exponent of $F\in\M^\times$. If $F$ is also a form i.e. an {\em inverse form} and  $d=|F|\leq0$ is its total degree,    we write $F=\sum_{j=d}^0F_{j,d-j}x^jz^{d-j}$.  {\em When $F$ is understood, we write $F_j$ for $F_{j,d-j}$} on the understanding that $F_j=0$ for $(j,d-j)\prec \ee(F)$.

 {\em Throughout the paper, $F\in\M^\times$ denotes a  typical non-zero inverse form  and $m=|F|\leq0$.}  
 We will use a restriction of the exponential valuation $\vv$ for inverse forms: the {\em order} of $F$ is $\vv=\vv(F)=\max\{j: |F|\leq j\leq 0, F_j\neq0\}$. 

 We will often use  $a\in\K$ to {\em augment a (non-zero) inverse form} $F\in\M^\times$ :   $a\sm F=ax^{m-1}+Fz^{-1}$, of total degree $m-1=|F|-1$\,. 
  For example, $a\sm z^m=a\,x^{m-1}+z^{m-1}$.  A form $F\in\M^\times$ defines  {\em inverse subforms} $\{F^{(i)}: m\leq i\leq \vv\}$ by $F^{(\vv)}=x^\vv$ and
  $$F^{(i)}=F_i\sm F^{(i+1)}=F_ix^i+F^{(i+1)}z^{-1}=F_ix^i+\cdots +x^\vv z^{i-\vv}\mbox{for \ }m\leq i\leq \vv-1.$$ 
 If $F_i\neq0$ then $\E(F^{(i)})=(i,0)$ for $m\leq i\leq \vv$. 
 
{\em The following inductive principle will often be used to prove a result for an arbitrary  inverse form}:
  
  (i)  prove the result for $x^\vv$ (the inductive basis)
  
  (ii) assuming the result for  $F$, let $a\in\K$ be arbitrary and   prove the result for $a\sm F$.

\subsection{The Module of Inverse Polynomials}
We recall the $\R$ module $\M=\K[x^{-1},z^{-1}]$ of { inverse polynomials}.
For  $x^i\in\R$ and $x^j\in\M$
\begin{eqnarray}\label{module}
x^i\circ x^j=\left\{\begin{array}{ll}
x^{i+j}&\mbox {if }x^{i+j}\in\M\\
0&\mbox{otherwise.}
\end{array}
\right.
\end{eqnarray}
The $\R$ module structure on $\M$ is obtained by linearly extending 
Equation (\ref{module})  to all of $\R$ and $\M$.   By linearity, we can without loss of generality assume that an inverse form $F$ satisfies $F_\vv=1$ i.e. $F=F_mx^m+\cdots +F_{\vv-1}x^{\vv-1}z^{m-\vv+1}+x^\vv z^{m-\vv}$. 
\bl \label{prebasic} (\cite[Lemma 3.1]{N15b}) If $\varphi\in\R^\times$, $F\in\M^\times$ are  forms  and $d=|\varphi|+|F|$ then (i) $$\varphi\circ F= \sum_{i=d}^0[\varphi\cdot F]_i\,x^i\,z^{d-i}$$

(ii) if $d>0$ then $\varphi\circ F=0$

(iii) if $\varphi\circ F\neq0$ then $\varphi\circ F$  is a form of total degree $d\leq0$. 
\el

\bl \label{f/z} (\cite[Lemma 3.8]{N15b}) We have (i) $z\circ  (a\sm F)= F$ (ii) if $\varphi\in z\,\R$ is a form then $\varphi\in\I_{ a\sm F}$ if and only if $\varphi/z\in \I_F$.
\el

   \subsection{The Ideal $\I_F$}
 Let $F$ be an inverse form. The {\em annihilator ideal} of $F$ is $\I_F=\{\varphi\in\R: \varphi\circ F=0\}.$

\bp (\cite[Proposition 3.7]{N15b}) \label{homog} The ideal $\I_F$ is homogeneous.
\ep

The next result will be our inductive basis.

\bp \label{001}((\cite[Proposition 3.8]{N15b}) If $F=x^m$ then  $\I_F=\langle x^{1-m},z\rangle$.
 \ep
We will also need the following elementary results.

\bd  \label{viable}  (\cite[Definition 3.13]{N15b}) {\rm  We will call  $f\in\R^2$   {\em a viable ordered pair  for}  $\I_F$  if $f_1,f_2$ are monic non-zero forms, $\I_F=\langle f_1,f_2\rangle$, $f_1\not\in\langle z\rangle$, $f_2\in\langle z\rangle$ and  $|f_1|+|f_2|=2-m$. }
\ed 

From Proposition \ref{001}, if $F=x^m$ then $(x^{1-m},z)$ is viable for  $\I_F$. 

\subsection{The  Integer $\lambda_F$}
The following definition makes sense since  $x^{1-|F|}\in\I_F$\,.
\bd (\cite[Definition 3.15]{N15b}) For an inverse form $F$, we define $\lambda_F\in\N$ by $\lambda_F=\min\{|\varphi|:\ \varphi\in\I_F\cap\Phi\}.$
 When $F$ is understood, we put  $\lambda_i=\lambda_{F^{(i)}}$ for $m\leq i\leq\vv$. 
\ed
\bp \label{Lambda}(\cite[Proposition 3.15]{N15b}) If $f$ is viable for $\I_F$ then  $\lambda_F=|f_1|$.
\ep
\section{The Construction}
We recall the main constructive result of \cite{N15b} used to obtain a viable ordered pair for $\I_{a\sm F}$ for one for $\I_F$.

\subsection{The Discrepancy}
The following characterisation improves \cite[Proposition 4.1]{N15b}.
\bp\label{Delta} Let  $\varphi\in\R^\times$ be a form. If $G= a\sm F$ and $d=|\varphi|+|G|\leq0$ then  $$\varphi\circ G=[\varphi\cdot G]_d\,x^d+(\varphi\circ F)z^{-1}.$$ In particular, $\varphi\in\I_G$ if and only if $[\varphi\cdot G]_d=0$ and $\varphi\in\I_F$. 
\ep
\bpr  From Lemma \ref{prebasic} 
\begin{eqnarray*}
\varphi\circ G&=&\sum_{d\leq i\leq 0} [\varphi\cdot G]_{i,d-i} \,x^i\,z^{d-i}
=[\varphi\cdot G]_d\,x^d+\sum_{d<i\leq0} [\varphi\cdot G]_{i,d-i}\, x^i\,z^{d-i}\\
&=&[\varphi\cdot G]_d\,x^d+\sum_{d<i\leq0} [\varphi\cdot (ax^{m-1}+Fz^{-1})]_{i,d-i}\, x^i\,z^{d-i}\\
&=&[\varphi\cdot G]_d\,x^d+a\sum_{d<i\leq0} [\varphi\cdot x^{m-1}]_{i,d-i}\,x^i\,z^{d-i}+\sum_{d<i\leq0}\,F_{i,d-i+1}\, x^i\,z^{d-i}\\
&=&[\varphi\cdot G]_d\,x^d+a\,S+T\end{eqnarray*}
say. Now $S=\sum_{d<i\leq0}\, \varphi_{i-m+1,d-i}\,x^i\,z^{d-i}=0$ since if $d-i< 0$ then  $\varphi_{i-m+1,d-i}=0$. Secondly,  let $e=|\varphi|\geq 0$ and $m=|F|$. Then $d<i$ if and only if $e+m\leq i$ and  again using Lemma \ref{prebasic} 
$$T=\left(\sum_{e+m\leq i\leq0}\,F_{i,e+m-i}\, x^i\,z^{e+m-i}\right)z^{-1}=(\varphi\circ F)\,z^{-1}.$$
In particular,  $[\varphi\cdot G]_d\,x^d$ cannot cancel with any term of $T=(\varphi\circ F)z^{-1}$ and this proves the necessity of the second statement.
\epr

First we discuss a 'discrepancy' which shows how $a$ and $\I_F$ affect $\I_{a\sm F}$. 
This is our analogue of  'discrepancy' as introduced in \cite{Ma69}.
\bd {\rm If     $\varphi\in\I_F^\times$ is a form and $G= a\sm F$ then the {\em discrepancy} of $\varphi$ and $G$ is
 $$\Delta(\varphi;G)=[\varphi\cdot G]_{|\varphi|+|G|}\in\K\mbox{\ \ \ \  if }|\varphi|+|G|\leq0$$ and $\Delta(\varphi;G)=0$ otherwise.}
\ed

Since $|f_i|+|G|\leq0$, $\Delta_i=[f_i\cdot G]_{|f_i|+|G|}$ for $i=1,2$. From Proposition \ref{Delta}, $\J\subseteq \I$.
 \subsection{The Inductive Step}
As in \cite{N15b}, for $d\in\Z$,  define $\ol{d},\ul{d}\in\Z$ by $\ol{d}=\max\{d,0\}$ and $\ul{d}=\min\{d,0\}$. 
Then  $\ol{d}+\ul{d}=d$. Secondly
\bd {\rm Let  $d=d_f=|f_2|-|f_1|$. If  $f_1,f_2\not\in\J$ let $q_f$ be the quotient $q=q_f=\Delta_1/\Delta_2$ and 
$$f_1\ominus f_2=f_1\ominus_{d,\,q} f_2=x^{\ol{d}}f_1-q\, x^{-\ul{d}}f_2\,.$$
It is convenient to put $f_1\ominus f_2=f_1$ if $\Delta_1=0$.}
\ed
We will omit $d_f,q_f$ when they are clear from the context. Note that $d_f$ is well-defined since if $f$ is viable for $\I$ then $f_1,f_2\neq 0$ and $q_f\in\K^\times$ is well-defined since $f_2\not\in\J$. Of course if $\K=\mathrm{GF}(2)$ then $q=1$ for any $f$.

\ul{For the remainder of the paper we will adopt the following notation:}

$F$ is a form in $\M^\times$, $a\in\K$ and $G=a\sm F$,  $\I=\I_F$ and $\J=\I_G$

$f$ is viable for $\I$, $d= |f_2|-|f_1|$ and $\Delta_i=\Delta(f_i; G)$ for $i=1,2$.

 $f_1$ is {\em active} if  $\Delta_1\neq 0$ and $d>0$, and {\em inactive} otherwise. 

\bt \label{template}(\cite[Proposition 4.6, Theorem 4.12]{N15b}) Suppose that  $f_2\not\in\J$\,. Then 

(i) 
if $g_1=f_1\ominus  f_2$, $g_2=f_2\,z$ if $f_1$ is inactive  and $g_2=f_1\,z$ otherwise, then $g$ is viable for $\J$ and 
$|g_1|=\max\{|f_1|,|f_2|\}=\max\{\lambda_F,2-m-\lambda_F\}$. 

(ii) $d_g=1-|d_f|$ and  for any $b\in\K$,  $q_g=\Delta(g_1;b\sm G)/\Delta_2$ if $f_1$ is inactive and $q_g=\Delta(g_1;b\sm G)/\Delta_1$ otherwise.\et

 Together with Proposition \ref{001}, we easily obtain \cite[Algorithm 4.22]{N15b} which does not use the last 'length change' or the variable $x=n-m$ of \cite{Ma69}.

\section{Simplifying the Groebner Basis Results}

\subsection{The Constructed Minimal Basis $\F$}
Let $I$ be a non-zero ideal of $\R$ and $G\subset \R^\times$ be finite. We recall  that  $G$ is a (grlex) {\Gb} of $I$ if for every $f\in I^\times$, there is a $g\in G$ with  $\LT(g)|\LT(f)$ (where leading monomials are with respect to grlex). Furthermore, a homogeneous ideal has a \Gb consisting of forms, \cite[Theorem 2, p. 380]{IVA}.
 Further, $G$ is said to be a {\em minimal \Gb} of $I$ if it is a \Gb for $I$ and for all $i$, $g_i$ is monic and  for $j\neq i$,  $\LT(g_i)\nmid \LT(g_j)$. 

\bd \label{templateFD} (\cite[Definition 5.2]{N15b} {\rm If $F=x^\vv$, we put $\F=\{x^{1-\vv},z\}$. Inductively, if $\F=\{f_1,\ldots,f_{|\F|}\}$ has been defined, define $\G=\{g_1,\ldots,g_{|\G|}\}$ by $g_1=f_1\ominus f_2$  and for $2\leq i\leq |\F|$
  $$g_i= \left\{\begin{array}{ll}
f_i\,z&\mbox{ if\ }f_1 \mbox{ is inactive}\\
 f_{i-1}\,z& \mbox{ otherwise.}\end{array}\right.
$$
}
\ed
An example $\F$ is given in Example \ref{ADex4} below. From \cite[Corollary 5.4]{N15b}, $|\F|\leq\lambda_F+1$.

\bt\label{minGb}For any  $F$, we can construct a  minimal \Gb $\F$ for $\I$.
\et
\bpr  We prove this by induction. If $F=x^\vv$, then by Proposition \ref{001}, $\F=\{x^{1-\vv},z\}$, a minimal \Gb for $\I$. For the inductive step, let $\F=\{f_1,\ldots,f_c\}$ be a minimal \Gb for $F$ and $\G$ be as defined above. Each element of $\G$ is monic by construction.  Let $g\in\J$. If $z|g$ then $g/z\in\I$ by Lemma \ref{f/z}. Since $\F$ is a \Gb for $\I$,  $\LT(f_i)|\LT(g/z)$ for some $i$. Then $\LT(g_i)=\LT(f_i)z|\LT(g)$. If on the other hand $z\nmid g$ then $|g|\geq \lambda_G=|g_1|$, so $\LT(g_1)|\LT(g)$.  
If $\G$ is not minimal then for some $i$ and $j\neq i$, $\LT(g_i)|\LT (g_j)$. If $i=1$ then $\LT(g_1)|\LT(g_j)$ for some $j\geq 2$. Since $\LT(f_1)|\LT(g_1)$ this implies that  $\LT(f_j)z=\LT(g_j)=t\,\LT(f_1)$ for some term $t$. Then $z|t$ as $z\nmid\LT(f_1)$ and  $\LT(f_j)=(t/z)\LT(f_1)$ so that $\LT(f_1)|\LT(f_j)$, which contradicts the minimality of $\F$. So we can assume that $i\geq 2$. If  $j=1$ then $z|\LT(g_1)$ which is impossible. Hence $i,j\geq 2$ and $\LT(f_i)z=\LT(g_i)$ divides $\LT(g_j)=\LT(f_j)z$ i.e. $\LT(f_j)|\LT(f_i)$ and $\F$ is not minimal. Therefore  $\G$ is a minimal \Gb for $\J$. This completes the inductive step and the proof.
\epr

It follows from Buchberger's Criterion (\cite[Theorem 6, p. 85]{IVA}) that  the  syzygy polynomial  $\mathrm{S}(f_i,f_j)=0$ for $1\leq i<j\leq|\F|$, as was shown in \cite[Subsections 5.3, 5.4]{N15b}.
 \subsection{Exponents and $\D_F$}

The following definition was motivated in \cite[Subsection 5.2]{N15b}.
 \bd \label{templateD} (\cite[Definition 5.7]{N15b} {\rm If $F=x^\vv$, we put  $\D=\mathcal{D}_F=(\vv,\vv+1)$ and $\lambda_{\D_2}=0$. Inductively, if  $\D$ has been defined, define $\mathcal{E}=\calE_G$ by $\mathcal{E}_1=m-1$  and for $2\leq i\leq |\F|$
  $$\mathcal{E}_i= \left\{\begin{array}{ll}
 \mathcal{D}_i&\mbox{ if\ }f_1 \mbox{ is inactive}\\
\mathcal{D}_{i-1}& \mbox{ otherwise.}\end{array}\right.
$$
}
\ed

For example, $\mathcal{E}=(m-1,\mathcal{D}_2,\ldots,\mathcal{D}_{|\F|})$ if $f_1$ is inactive, $\mathcal{E}=(m-1,\mathcal{D}_1,\ldots,\mathcal{D}_{|\F|})$ otherwise, and $|\D|=|\F|$. We include an example for the convenience of the reader.
\be \label{ADex4} (\cite[Example 5.6]{N15b}) {\rm For $\K=\mathrm{GF}(2)$ and $F=x^{-6}z^{-1}+x^{-4}z^{-3}+x^{-3}z^{-4}+z^{-7}$}
\begin{center}
\begin{tabular}{|r|l|r|}\hline
$m$     &$f$ & $\D$ \\\hline\hline
$0$   & $(x,z)$  & $(0,1)$ \\\hline
 $-1$ & $(x,z^2)$ & $(-1,1)$  \\\hline
  $-2$ &$(x,z^3)$ & $(-2,1)$ \\\hline
$-3$  & $(x^3+z^3,xz,z^4)$   & $(-3,-2,1)$ \\\hline
$-4$  & $(x^3+x^2z+z^3,xz^2,z^5)$  & $(-4,-2,1)$\\\hline
$-5$     & $(x^3+x^2z+xz^2+z^3,xz^3,z^6)$  & $(-5,-2,1)$\\\hline
$-6$   & $(x^4+x^3z+x^2z^2,(x^3+x^2z+xz^2+z^3)z,xz^4,z^7)$ & $(-6,-5,-2,1)$ \\\hline
$-7$    & $(x^4+xz^3+z^4,(x^3+x^2z+xz^2+z^3)z^2,xz^5,z^8)$ & $(-7,-5,-2,1)$. \\\hline
\end{tabular}
\end{center}
\begin{figure}\label{fig3}
\caption{$\ee(\F)$ and  $\D=(-7,-5,-2,1)$ for Example \ref{ADex4}.}
 \begin{center}
\setlength{\unitlength}{0.75cm}
\begin{picture}(1,9)
\thicklines
\put(0,9){\line(0,-9){9}}
\put(0,0){\line(5,0){5}}
\put(4,1){\circle{0.2}}
\put(3,3){\circle{0.2}}
\put(3,4){\circle{0.2}}
\put(2,5){\circle{0.2}}
\put(1,6){\circle{0.2}}
\put(1,7){\circle{0.2}}
\put(0.05,8.05){$^{(0,8)=\ee(\F_4)=(\lambda_{\D_4},\D_4-m)}$}
\put(1.05,5.05){$^{(1,5)=\ee(\F_3)=(\lambda_{\D_3},\D_3-m)}$}
\put(3.05,2.05){$^{(3,2)=\ee(\F_2)=(\lambda_{\D_2},\D_2-m)}$}
\put(4.05,0.05){$^{(4,0)=\ee(\F_1)=(\lambda_{\D_1},\D_1-m)}$}

\put(0,8){\line(1,0){1}}
\put(0,8){\circle*{0.25}}
\put(1,8){\line(0,-3){3}}
\put(1,5){\line(2,0){2}}
\put(1,5){\circle*{0.25}}
\put(3,5){\line(0,-3){3}}
\put(3,2){\line(1,0){1}}
\put(3,2){\circle*{0.25}}
\put(4,2){\line(0,-2){2}}
\put(4,0){\circle*{0.25}} 
\end{picture}
\end{center}
\end{figure}

  \eex
The following lemma  generalises Theorem 4.10 and Proposition 5.9 of \cite{N15b}.
\bl \label{exp} If $c=|\F|$ then

(i) $\D_1<\cdots<\D_c$ 
 and  for $2\leq i\leq c$
$$\lambda_{\D_{i-1}}=\lambda_{\D_{i-1}+1}=\cdots=\lambda_{\D_i-1}>\lambda_{\D_i}$$

(ii) $\ee(f_i)=(\lambda_{\D_i},\D_i-m)$ for $1\leq i\leq c$ and  $\LT(f_{i-1})>_{\mathrm{lex}}\LT(f_i)$ for $2\leq i\leq c$

(iii) $\lambda_{\D_{i-1}}+\lambda_{\D_i}=2-\D_i$ for $2\leq i\leq c$.
 \el
\bpr {\em Let $F=x^\vv$.} Then from Proposition \ref{001}, $\F=(x^{1-\vv},z)$ and $c=2$ so that 
 
 (i) $\D_1=\vv<1+\vv=\D_2$ and $\lambda_{\D_1}=1-\vv\geq1>0=\lambda_{\D_2}$. 
 
 (ii) We have $\ee(f_1)=(1-\vv,0)=(\lambda_{\D_1},\D_1-\vv)$ and $\ee(f_2)=(0,1)=(\lambda_{\D_2},\D_2-\vv)$. 
 
 (iii)  $\lambda_{\D_1}+\lambda_{\D_{2}}=1-\vv+0=2-\D_2$.
 
Suppose that the result is true for $m$ and let $n=m-1$.   {\em If $f_1$ is inactive} then $|\G|=c$, $\G= \{f_1,f_2\,z,\ldots,f_c\,z\}$ and $\calE=(n,\D_2,\ldots,\D_c)$.

(i) From the definitions, $\calE_1=n<m=\D_1<\D_2=\calE_2<\cdots<\D_c=\calE_{|\G|}$ and 
$\lambda_{\calE_{i-1}}=\lambda_{\calE_{i-1}+1}=\cdots=\lambda_{\calE_i-1}>\lambda_{\calE_i}$ for $2\leq i\leq c$ follows immediately from the inductive hypothesis. 

(ii) $\ee(g_1)=\ee(f_1)=(\lambda_m,0)=(\lambda_n,0)=(\lambda_{\calE_1},\calE_1-n)$ and for $2\leq i\leq c$
$$\ee(g_i)=\ee(f_i)+(0,1)=(\lambda_{\D_i},\D_i-m)+(0,1)=(\lambda_{\calE_i},\calE_i-n).$$ 

(iii) For $2\leq i\leq c=|\G|$,  $\lambda_{\calE_{i-1}}+\lambda_{\calE_i}=\lambda_{\D_{i-1}}+\lambda_{\D_i}=2-\D_i=2-\calE_i$.

{\em Finally suppose that $f_1$ is active}. Then $c=|\G|+1$, $\G= \{f_1\ominus f_2\,,f_1\,z,\ldots,f_c\,z\}$ and $\calE=(n,\D_1,\ldots,\D_c)$.

(i) $\calE_1=n<m=\D_1=\calE_2<\D_2=\calE_3<\cdots<\D_c=\calE_{|\G|}$ and $\lambda_{\D_{i-1}}=\lambda_{\D_{i-1}+1}=\cdots=\lambda_{\D_i-1}>\lambda_{\D_i}$ for $2\leq i\leq c$. Now $\calE_1+1=\D_1=\calE_2$ so $\lambda_{\calE_1}>\lambda_{\D_1}=\lambda_{\calE_2}$ and for $3\leq i\leq c+1$,
$$\lambda_{\calE_{i-1}}=\lambda_{\calE_{i-1}+1}=\cdots=\lambda_{\calE_i-1}>\lambda_{\calE_i}$$
follows immediately from the inductive hypothesis. 

(ii) $\ee(g_1)=\ee(f_1\ominus f_2)=(\lambda_m+d,0)=(\lambda_n,0)=(\lambda_{\calE_1},\calE_1-n)$ and for $2\leq i\leq c+1$ 
$$\ee(g_i)=\ee(f_{i-1})+(0,1)=(\lambda_{\D_{i-1}},\D_{i-1}-m)+(0,1)=(\lambda_{\calE_i},\calE_i-n).$$

(iii) Since $f$ is viable for $F$, $|f_1|+|f_2|=2-m$ and $$\lambda_{\calE_1}+\lambda_{\calE_2}=(d+\lambda_{\D_1})+\lambda_{\D_1}=
|f_2|-|f_1|+2\lambda_F=|f_2|+|f_1|=2-m=2-\calE_2$$
 and for $i\geq 3$,
$\lambda_{\calE_{i-1}}+\lambda_{\calE_i}=\lambda_{\D_{i-2}}+\lambda_{\D_{i-1}}=2-\D_{i-1}=2-\calE_i$.
\epr

\brs {\em Initially $d=\vv\leq0$ and $\lambda_{\D_1}=1-\vv$. From (i)  each run of identical $\lambda_i$ is followed by a single increase, and $\lambda_{\D_c},\ldots,\lambda_{\D_1}$ is the 'linear complexity profile' of $F$. 

From (ii), since $\ee(f_i)=(\lambda_{\D_i},\D_i-m)$ and $\lambda_{\D_{i-1}}>\lambda_{\D_i}$ for $2\leq i\leq |\F|$, $\F$ is naturally lex-ordered i.e. we can write $\F=(f_1,\ldots,f_c)$ in strictly decreasing lex order. Secondly, $|f_i|=\lambda_{\D_i}+\D_i-m$ for $1\leq i\leq |\F|$.}
\ers

Lemma \ref{exp} gives a direct proof of the following:
\bc ( \cite[Corollary 5.18]{N15b}) $\dim_\K(\R/\I)=|f_1||f_2|$.
\ec
\bpr Since $\F$ is a minimal \Gb for $\I$, $\langle\LT(\I)\rangle=\langle\{\LT(f_i):1\leq i\leq |\F|\}\rangle$ and so the $\K$ vector space $\R/\I$ is spanned by $\{x^j: x^j\not \in\langle\LT(f_i)\rangle$ by \cite[Proposition 4]{IVA}. Also $\dim_\K(\R/\I)$ is the sum of the areas of the rectangles with lengths  $(\D_i-\D_{i-1})$ and heights $\lambda_{\D_i}$ for $2\leq i\leq |\F|$ i.e. 
$\dim_\K(\R/\I)=\sum_{i=2}^{|\F|} (\D_i-\D_{i-1})\,\lambda_{\D_{i-1}}$. By Lemma \ref{exp}, this sum is $\sum_{i=0}^m \lambda_i=|f_1||f_2|$ by \cite[Proposition 4, p. 232]{N15b}.
\epr
\subsection{The Reduced \Gb $\ol{\F}$}\label{RGbsub}
  In this subsection we show how to modify the minimal Groebner basis  $\F$ to obtain  the reduced Gb (RGb) of $\I$ without using the notion of an essential form. 
Recall that (i) a minimal \Gb $\mathsf{G}$ is {\em reduced} if for all $\mathsf{g}\in\mathsf{G}$, no monomial of $\mathsf{g}$ is in $\langle \LT(\mathsf{G}\setminus\mathsf{g})\rangle$ and (ii) 
 $\langle \mathsf{G}\rangle$ has a unique (grlex) RGb,  which we write as $\ol{\mathsf{G}}$. The standard method for obtaining $\ol{\mathsf{G}}$ is to successively replace $\mathsf{G}$ by $(\rem_{\mathsf{G}\setminus\mathsf{g}}\,\mathsf{g})\cup(\mathsf{G}\setminus\mathsf{g})$,  \cite[p. 92]{IVA}.  We say that a monomial $M$ {\em can be reduced by} $\mathsf{G}$ or $\mathsf{G}$ {\em reduces} $M$ if $\LM(\mathsf{g})|M$ for some $\mathsf{g}\in\mathsf{G}$.
We can always construct the RGb $\ol{\F}$ of $\I$ according to the  method of \cite{IVA}. 

We will see that the standard method can be considerably improved in  our case: it will suffice to replace $f_1$ by $\rem_{f_2}\,f_1$ if $\F$ is not reduced. 
  \bl\label{j=2}(\cite[Lemma 5.19]{N15b}) 
 If  $f_i$ reduces a monomial $M$ of $f_1$  then $i=2$. 
 \el
{\em Proof.} We cannot have $i=1$ and if $|\F|=2$ there is nothing to prove, so let $i\geq 3$. We will derive a contradiction. If $\D$ is the degree tuple then $\LM(f_i)=x^{\lambda_{\D_i}}z^{\D_i-m}$ by Lemma \ref{exp}.  In particular, $m<\D_i$ and $z|\LM(f_i)$. Since $\LM(f_i)|M$, $z|M$ and we can write $M=x^{\lambda_F-p}z^p$ for some $p$ such that $1\leq p\leq \lambda_F$ and $\lambda_{\D_i}\leq \lambda_F-p$.   From Lemma \ref{exp} we also have $\lambda_F=2-\D_2-\lambda_{\D_2}$. Since $i\geq3$, $\D_2\leq\D_i-1$ and $\lambda$ is non-decreasing, $\lambda_{\D_2}\geq\lambda_{\D_i-1}$. This gives

$$\lambda_{\D_i}\leq\lambda_F-p=(2-\D_2-\lambda_{\D_2})-p
\leq 2-\D_2-\lambda_{\D_i-1}-p.$$
From Lemma \ref{exp} again $\lambda_{\D_i-1}=\lambda_{\D_{i-1}}=2-\D_i-\lambda_{\D_i}$\,, so 
$\lambda_{\D_i}\leq2-\D_2-(2-\D_i-\lambda_{\D_i})-p=-\D_2+\lambda_{\D_i}-p+\D_i$.
Finally $-\D_2<-\D_1=-m$ and  $\D_i-m\leq p$ since $\LM(f_i)|M$. Combining these inequalities, $i\geq 3$ gives the contradiction 
$$\lambda_{\D_i}\leq-\D_2+\lambda_{\D_i}-p+\D_i<-m+\lambda_{\D_i}-p+\D_i=\lambda_{\D_i}+(\D_i-m-p)\leq\lambda_{\D_i}\,.\ \square$$

\bt \label{RGb} (Inductive RGb) Assume that $\F$ is reduced. If $g$ is viable for $\J$ and $\G$ is the form tuple for $G$ then 

(i)  if $\Delta_1=0$  then $\G$ is reduced 

(ii)  $\G$  is  reduced if and only if $g_1=\rem_{g_2}\,g_1$

(iii)  if $|g_2|> |g_1|$ then $\G$ is reduced.
\et
\bpr  Let $\Delta_1$ and $c=|\F|$. (i) If  $f_1$ is inactive then $\G=(f_1,f_2z,\ldots,f_{c}z)$. Since $\langle \LM(f_i)\,z:i>1\rangle\subset\langle \LM(f_i):i>1\rangle$ and $\F$ is reduced, a monomial of $g_1=f_1$ cannot be reduced.  Likewise, if $2\leq i\leq |\G|=|\F|$\,,  $\langle \LM(f_1),\{\LM(f_j)z: j\neq i\}\rangle\subset\langle \LM(f_1),\LM(f_j): j\neq i\}\rangle$ so no monomial of $g_i=f_iz$ can be reduced.

 (ii) It suffices to show that if $\G$ is not reduced then a monomial of $g_1$ lies in $\langle\LT(g_2)\rangle$.  From (i) we have $\Delta_1\neq 0$, so either (a) $d\leq 0$, $\LM(g_1)=\LM(f_1)$  and $\G=(g_1,f_2z,\ldots,f_{c}z)$ or (b) $d>0$,  $\G=(g_1,f_1z,\ldots,f_{c}z)$ and $\LM(g_1)=x^d\,\LM(f_1)$. As before, no term of $f_i\,z$ lies in  $\langle \LM(f_j)z: j\neq i\rangle$ since $\F$ is reduced. Suppose that for some $i$, $2\leq i\leq |\G|$\,, we could reduce a term of $f_i\,z$  by $\LT(g_1)$. Since $\R$ has unique factorisation and $z\nmid g_1$, a term of $f_i$ would be reducible by either $\LM(g_1)=\LM(f_1)$ or by $x^d\,\LM(f_1)$ i.e. be reducible by $\LM(f_1)$ and so $\F$ would not be reduced. Thus if $\G$ is not reduced, a monomial $M$ of $g_1$ can be reduced by $\G_j$ for some $j>1$. If $|\G|=2$ we are done and if $|\G|\geq 3$ then Lemma \ref{j=2} implies that $j=2$.

(iii) From (ii) if $\G$ is not reduced then $g_1$ can be reduced by $g_2$ and so  $|g_2|\leq |g_1|$.
\epr

\bc\label{RGbT} For any  $F$ we can construct  the unique RGb $\ol{\F}$ of $\I$\,. 
\ec
\bpr  If $F=x^\vv$ then $\F=(x^{1-\vv},z)$ is reduced. Suppose inductively that   $\F$   is reduced. If $\Delta_1=0$ then $\G$ is already reduced by Theorem \ref{RGb}, so assume that $\Delta_1\neq 0$. If $\G$ is not reduced then $d\leq 0$ and we replace $g_1$ by  $\rem_{g_2}\,g_1$ as per Theorem \ref{RGb} so that $\G$ will be  reduced.  Finally,  $(\rem_{g_2}\,g_1,g_2)$ is viable for $\J$ and a suitable input for Theorem \ref{template} (which does not require that  the input viable pair be  constructed via  Theorem \ref{template}).This completes the induction. 
\epr

For completeness we include the next corollary which strengthens \cite[Corollary 4.36]{N15b}.
\bc (Cf. \cite[Lemma 7]{AD}) If    $\lambda_F\leq 1-m/2$ then $\F$ is reduced.
\ec
\bpr  We show that if $\F$ is not reduced then $2\lambda_F\geq2-m$. By  Lemma \ref{j=2}, if $\F$ is not reduced then $f_1$ can be reduced by $f_2$ and hence $|f_1|\geq |f_2|$. Proposition \ref{Lambda} now implies that $ \lambda_F=|f_1|\geq |f_2|=2-m-|f_1|=2-m-\lambda_F$.
\epr

\section{Additional Simplifications}

For good measure we reprove some additional results in a simpler way.
\subsection{A Characterisation} 

The proof of (c) $\Rightarrow$ (d) in \cite[Corollary 4.36]{N15b} relied on the factorisation of $f_2$ in Theorem  4.33, {\em loc. cit.} and in particular the technical notion of a 'triple' for an 'essential' $F$. 
We give a direct proof which does not rely on the triple of an essential inverse form.

\bc (\cite[Corollary 4.36]{N15b}) The following are equivalent:
\begin{tabbing}
\hspace{1cm}\= (a) $\lambda_F<1-\frac{m}{2}$\\
\>(b) $|f_1|<|f_2|$\\
\>(c) $f_1$ is unique: if $\varphi\in\I\cap\Phi$ and $|\varphi|=|f_1|$ then $\varphi=f_1$\\
\>(d)  $f_2$ is not unique: there is a monic form $\varphi\in\I^\times$ with  $z|\varphi$, $|\varphi|=|f_2|$ and $\varphi\neq f_2$.
\end{tabbing}
\ec
\bpr We prove that  (c) $\Rightarrow$ (d) only (the other parts are proved without appealing to \cite[Theorem 4.33, Proposition 4.35]{N15b}). As in \cite{N15b}, (b) $\Rightarrow$ (c), so $d>0$ and hence $m<0$. We claim that $\varphi=f_1z^d-f_2$ satisfies the properties of (d). Firstly, if $\varphi=0$ then $x^{\lambda_F}|f_2$ and $\lambda_2\geq \lambda_F$, which contradicts Lemma \ref{exp} for $i=1$. Thus $\varphi\neq0$ and $|\varphi|=|f_1|+d=|f_2|$ i.e. $\varphi$ is a form. Since $z|f_2$ and $d>0$, $z|\varphi$. Finally, Lemma \ref{exp} implies $\LT(f_1)>_{\mathrm{lex}} \LT(f_2)$ so that $\LT(\varphi)=\LT(f_1)z^d$ and $\varphi$ is monic.
\epr

\subsection{Factorising the Forms of $\F$}
We reprove Theorem 4.33 and Proposition 4.35 of \cite{N15b} without using triples or the factorisations $f_i=f_i'\,z^p$ where $z\nmid f_i'$. Again, treating the cases $F=1$, $F$ geometric and $F$ 'essential' separately is no longer necessary. 

\bl (\cite[Theorem 5.10]{N15b})  For $1\leq i\leq |\F|$,    $f_i=f_i^{(\D_i)}\,z^{\D_i-m}$ where $f_i^{(\D_i)}=f_i/z^{\D_i-m}$ and $z\nmid f_i^{(\D_i)}$.
\el
\bpr  From Lemma \ref{exp}, $\LM(f_i)=x^{\lambda_{\D_i}}z^{\D_i-m}$ for $1\leq  i\leq |\F|$. The non-leading terms of $f_i$ have powers of $z$ at least $\D_i-m$, so $z^{\D_i-m}$ divides $f_i$ and no larger power of $z$ can divide $f_i$.  In other words, we can take $f_i^{(\D_i)}=f_i/z^{\D_i-m}$.
\epr

Putting $i=2$ in the following result is our analogue of  \cite[Equation (11), p.123]{Ma69}.
\bp  \label{D_i}For $2\leq i\leq |\F|-1$, $\D_i=\min\{j: \lambda_{\D_{i-1}}>\lambda_j\}$.
\ep
\bpr Immediate from Lemma \ref{exp}. Or induct.
\epr

\subsection{Finite Sequences}
We give a simpler justification of \cite[Algorithm 6.14]{N15b}, an analogue of the BM algorithm.  This derivation does not use the separate cases, triples, factorisation, the last 'length change'   of \cite[Formula (11)]{Ma69}  nor the variable '$x=n-m$' of the BM algorithm of {\em loc. cit}.

We write $s=s_0,\ldots,s_{n-1}$ for a  typical non-trivial finite sequence over $\K$ with $n\geq 1$ terms and $\vv=\min\{i: s_i\neq0\}$. As before, we can assume that $s_{\vv}=1$. We regard $s$ as being obtained from $0,\ldots,0,s_\vv$  by successively adding  the terms $s_{\vv+1},\ldots,s_{n-1}$.  

The inverse form of $s$ is 
$ F^{(s)}=\sum_{i=1-n}^0 s_{-i}x^iz^{1-n-i}\in\M^\times$
so that $\vv=-\vv(F^{(s)})$\,. We write $\I_s$ for the  annihilator ideal  $\I_{ F^{(s)}}$.  

 Recall that the {\em dehomogenisation} of $\varphi\in\R^\times$ is $\varphi^\vee(x)=\varphi(x,1)$ and if $z\nmid \varphi$ then   $\varphi_i=\varphi^\vee_i$ for $0\leq  i\leq|\varphi|=|\varphi^\vee|$.
Hence $\varphi\in\I_s\cap\Phi$ if and only if $[\varphi^\vee\cdot F^{(s)}]_i=0\mbox{ for }|\varphi|+1-n\leq i\leq 0$ or $\varphi^\vee$ is a (non-zero) {\em annihilating polynomial} of $s$. 
We call a non-zero annihilating polynomial of $s$ of least degree a {\em minimal polynomial (MP) of $s$}.  If $f$ is viable for $\I_s$ then $\mu_1=f_1^\vee$ is an MP of $s$. We will use an auxiliary polynomial $\mu_2=f_2^\vee$. Note that $\mu_1\ominus \mu_2=(f_1\ominus f_2)^\vee$ is also an annihilating polynomial of $s$.  

 We use the variable $d=|f_2|-|f_1|$ {\em as is}. Next we evaluate $\Delta_1=\Delta(f_1;  F^{(s_0,\ldots,s_i)})$:
\begin{eqnarray*}\Delta_1
=[f_1\cdot \sum_{j=-i}^0 s_{-j}\,x^jz^{-i-j}]_{|f_1|-i}=\sum_{j=0}^{|f_1|}\, [f_1]_j\, s_{j+i-|f_1|}=\sum_{j=0}^{|\mu_1|}\, [\mu_1]_j\, s_{j+i-|\mu_1|}.
\end{eqnarray*}

\br {\em The {\em linear complexity}  of $s$ is $|\mu_1|=\lambda_{F^{(s)}}$ from \cite[Proposition 6.7]{N15b}.
From Lemma \ref{exp}, $|\mu_2|=\lambda_{\D_2}$ and $|\mu_1|+|\mu_2|=2-\D_2$\,,  our analogue of
\cite[Formula (13)]{Ma69}. }

\er
Thus  (i) replacing $i$ by $-i$ (ii) dehomogenising $f_1,f_2$ (iii) evaluating $\Delta_1$ (iv) suppressing the statement $f_2\la f_2\,z$  in \cite[Algorithm 6.14]{N15b} justify the following algorithm.
 \begin{algorithm}\label{calPAs}(\cite[Algorithm 4.22]{N15b} modified for  a sequence, cf. \cite[Algorithm p. 124]{Ma69})
\begin{tabbing}
\noindent {\tt Input}: \ \ \=  Non-trivial sequence $s=s_0,\ldots,s_{n-1}$.\\

\noindent {\tt Output}: \> Minimal polynomial $\mu_1$ for $s$.\\\\

$\lceil$ $i\la -1$; {\tt repeat}   $i\la i+1$; {\tt until} $(s_i\neq0)$ $\vv\la i$; (* find $\vv$ *)\\\\
 $ \mu\la (x^{\vv+1},1)$; $\Delta_2\la 1$;\ $d \la  -\vv$; (* initialise *)\\\\

{\tt for} \= $i\la \vv+1$ {\tt to }$n-1$ {\tt do}\\\\

   \>$\lceil \ \Delta_1\la \sum_{j=0}^{|\mu_1|}\, [\mu_1]_j\, s_{j+i-|\mu_1|}\,; \ q\la \Delta_1/\Delta_2$;\\\\    
   \>{\tt if }$(\Delta_1\neq 0)$ {\tt then} \={\tt if} \=$(d\leq 0)$\ \={\tt then}\, $ \mu_1\la \mu_1-q\,  x^{-d}\, \mu_2 $\\\\
  \> \> \>\> {\tt else}
   $\lceil$\= $\psi\la \mu_1$; $\mu_1 \la  x^d\, \mu_1-q\, \mu_2$; $\mu_2\la \psi$; \\\\
   \>\> \> \>\>$\Delta_2\la  \Delta_1$; $d \la  -d;\rfloor$\\\\
   \> $d  \la  d+1; \rfloor$\\\\
{\tt return }$\mu_1.\rfloor$
\end{tabbing} 
  \end{algorithm}
  
 Finally, we have excluded the all-zero sequence as it complicates the algorithm for no conceptual gain; indeed we can simply set $\mu_1=1$ in this case.
 
   \section{The Maximal Regular Sequence}

As an addendum to \cite{N15b},we show that $f_1,f_2$ of Theorem \ref{template} is a maximal $\R$ regular sequence.  In  \cite[Theorem 1]{AD}, the authors show their  annihilator  is primary with associated prime $\m$. We begin with a different proof of this fact; in particular, the radical of $\I$ is $\m$. We use \cite{Sharp} as a general reference for commutative algebra.

\bl  \label{power} We have 
$\m^{1-m}\subseteq  \I \subseteq \m$; in particular, $\I$ is $\m$ primary and $\sqrt{\I}=\m$.
\el
\bpr   The ideal $\m^{1-m}$ is generated by the set $\{ x^i\,z^{1-m-i}\ :\ 0\leq i\leq 1-m\}$   and if $0\leq i\leq 1-m$ then
$$x^i\,z^{1-m-i}\circ F= \sum_{m\leq j\leq0}F_{j,m-j}\,  x^i\,z^{1-m-i}\circ x^j\,z^{m-j}.$$ If  $i+j\leq 0$  and $1-i-j\leq 0$ for some $i,j$ then  $1\leq i+j\leq 0$. Hence for all $i,j$ either $i+j\geq 1$  or $1-i-j\geq 1$ i.e.  the right-hand side vanishes and   $x^i\,z^{1-m-i}\in \I$. Thus $\m^{1-m}\subseteq  \I$. 

If  $F=x^\vv$ then by Proposition \ref{001} we have $\I=\langle x^{1-\vv},z\rangle\subseteq \m$. Suppose inductively that $m\leq i \leq \vv-1$ and $\I_{F^{(i+1)}}\subseteq \m$. From Lemma \ref{Delta}, $\I_{F^{(i)}}\subset \I_{F^{(i+1)}}$ and so by the inductive hypothesis, $\I_{F^{(i)}}\subseteq \m$. The remaining conclusions are standard.
\epr

\bc $(\R/\I,\m/\I,\K)$ is an Artinian local ring of dimension zero. 
\ec
\bpr A maximal ideal of $\R/\I$ is $\m'/\I$ where $\m'$ is a maximal ideal of $\R$ containing $\I$.  As $\m'$ is prime and $\sqrt{\I}=\m$, we have $\m\subseteq  \m'$ and since  $\m$ is maximal, $\m=\m'$.   Finally  $(\R/\I)/(\m/\I)\cong \R/\m\cong\K$. Lemma \ref{power} implies that  $(\m/\I)^{1-m}=0$ and hence the Noetherian ring $\R/\I$ has dimension zero and is Artinian.
\epr

 If $M$ is an $\R$ module, we write $\Zdiv(M)$ for the  zero divisors of $\R$ on $M$.
 Recall that  $g_1,g_2\in\R$ is an ($\R$) {\em regular sequence} (for $I=\langle g_1,g_2\rangle$) if $I$ is a proper ideal of $\R$, $g_1\neq0$   and $g_2\not\in\Zdiv(\R/\langle g_1\rangle)$, \cite[p. 312]{Sharp}.  Also, for  a proper ideal $I$ of $\R$, the common length of a maximal  regular sequence generating  $I$ is called $\mathrm{grade}(I)$.  

 The following is a special case of \cite[Lemma 16.7]{Sharp}.
 \bl \label{transp} Let $g_1,g_2\in\R^\times$ and $\langle g_1,g_2\rangle$ be a proper ideal of $\R$. If $g_2\not\in\Zdiv(\R/\langle g_1\rangle)$ then $g_1\not\in\Zdiv(\R/\langle g_2\rangle)$. In particular, $g_1,g_2$ is regular if and only if $g_2,g_1$ is.
 \el

\bt\label{regseq} The sequence  $f_1,f_2$ is  maximal regular  and $\mathrm{grade}(\I)=\mathrm{height}(\I)=2$. 
\et
\bpr  Since $F$ is non-zero, $\I$ is a proper ideal of $\R$ and  $f_1,f_2\in\R^\times$. We proceed by induction. Recall that $z\nmid f_1$ and $z|f_2$.
First the inductive basis: if $F=x^\vv$, then $f=(x^{1-\vv},z)$ by Proposition \ref{001}. If $f_2\in\Zdiv(\R/\langle f_1\rangle)$ i.e. for some $\alpha\not\in\langle x^{1-\vv}\rangle$ and $\beta\in \R$ we have $z\,\alpha= x^{1-\vv}\,\beta$ then since $z|\beta$, $\alpha= x^{1-\vv}\,\beta/z\in\langle x^{1-\vv}\rangle$  for a contradiction. Thus $x^{1-\vv},z$ is a regular sequence.

Suppose inductively that   $f_1,f_2$ is a regular sequence. Let  $g$ be viable for $\J$ as in Theorem \ref{template}. We will show that  $g_1\not\in\Zdiv(\R/\langle g_2\rangle)$, which suffices by Lemma \ref{transp}. 
Suppose  that $g_1\in\Zdiv(\R/\langle g_2\rangle)$ i.e. for some $\alpha\not\in\langle g_2\rangle$ and $\beta\in\R$ we have 
 $g_1\,\alpha= g_2\,\beta$.  If $f_1$ is inactive then  $g=(f_1-q\,x^{-d}f_2,f_2\,z)$  and if $g_1\in\Zdiv(\R/\langle g_2\rangle)$ then 
 $ (f_1-q\,x^{-d}f_2)\,\alpha=f_2\,z\,\beta$
where $\alpha\not\in\langle f_2\,z\rangle$. This implies that  $z|\alpha$, $ f_1\,\alpha/z=f_2(q\,x^d\,\alpha/z+\beta)$  where $\alpha/z\not\in\langle f_2\rangle$ i.e.  $f_1\in\Zdiv(\R/\langle f_2\rangle)$ which contradicts the inductive hypothesis by Lemma \ref{transp}.
 (ii)  If $f_1$ is active then $g=(x^d\,f_1-q\,f_2,f_1\,z)$ and 
  if $g_1\in\Zdiv(\R/\langle g_2\rangle)$ then 
 $ (x^df_1-q\,f_2)\,\alpha=f_1\,z\,\beta$
  where $\alpha\not\in\langle g_2\rangle$. This gives $f_2\,q\,\alpha/z=f_1(x^d\,\alpha/z-\beta)$ where $q\,\alpha/z\not\in\langle f_1\rangle$, so $f_2\in\Zdiv(\R/\langle f_1\rangle)$ again contradicting the inductive hypothesis via Lemma \ref{transp}.  Thus  $g_1\not\in\Zdiv(\R/\langle g_2\rangle)$ and $g_1,g_2$ is  regular, which completes the induction.

From Lemma \ref{power} and \cite[Proposition 16.22]{Sharp}, $\mathrm{grade}(\I)=\mathrm{grade}(\sqrt{I_F})=\mathrm{grade}(\m)$.
 Now $\K$  is Cohen-Macaulay since it is Artinian and hence so is $\R$ (see \cite[Theorem 17.33]{Sharp}) so by definition $\mathrm{grade}(\m)=\mathrm{height}(\m)$, and $\mathrm{height}(\m)=2$ by \cite[Corollary 14.33]{Sharp}. Thus $2=\mathrm{grade}(\I)=\mathrm{height}(\I)$  since $\R$ is Cohen-Macaulay. If the sequence $f_1,f_2$ were  not  maximal, we may extend it to a maximal  regular sequence of length strictly greater than 2 by \cite[Corollary 16.10]{Sharp}, contradicting $\mathrm{grade}(\I)=2$. 
\epr

\end{document}